\documentclass[11pt, a4paper]{article}
\usepackage{jcappub}
\usepackage{bm}
\usepackage{graphicx}
\usepackage{epstopdf}
\usepackage{hyperref}

\newcommand{\dpq}{\delta p_{rq}}

\newcommand{\dfq}{\delta \phi_q}
\newcommand{\drq}{\delta \rho_{rq}}
\newcommand{\duq}{\delta u_{rq}}
\newcommand{\ddfq}{\delta \dot{\phi}_q}

\newcommand{\tRH}{t_{\rm RH}}

\newcommand{\TRH}{T_{\rm RH}}
\newcommand{\aRH}{a_{\rm RH}}

\newcommand{\vv}{{\bf v}}
\newcommand{\n}{{\bf \nabla}}
\newcommand{\PP}{{\bf \Pi}}
\newcommand{\bP}{{\bf P}_q}

\title{Cosmological perturbations for an inflaton field coupled to radiation}

\author{Luca Visinelli}
\affiliation{Department of Physics and Astronomy, University of Utah, 115 South 1400 East \#201, Salt Lake City, Utah 84112-0830, USA}
\emailAdd{Luca.Visinelli@utah.edu}

\abstract{Within the framework of the interacting fluid formalism, we provide the numerical solution to the Boltzmann equation describing the evolution of an inflaton field coupled to radiation. We study the behavior of the system during the slow-roll regime, in the case in which an additional stochastic source term is included in the set of equations, and we recover the expression for the cosmological perturbations previously obtained in the Warm inflation scenarios.}

\arxivnumber{}

\begin{document}
\maketitle

\section{Introduction}

The mechanism of inflation~\cite{kazanas, starobinsky, guth, sato, albrecht, linde} addresses the problems of the flatness, homogeneity, and the lack of relic monopoles raised by the standard Big-Bang cosmology \cite{linde_book, kolb_book, weinberg_book}, and provides an explanation to the inhomogeneities observed in the Cosmic Microwave Background Radiation (CMBR)~\cite{mukhanov, guth_pi, hawking, starobinsky1, bardeen1}. It is usually postulated that during the inflationary period, the expansion of the Universe is driven by the potential energy of the so-called ``inflaton'' field. When the kinetic energy of the inflaton is no longer negligible with respect to its potential energy, the inflaton decays into lighter particles, which might be both hypothetical and Standard Model particles, and the Universe reheats and transits to a radiation-dominated epoch~\cite{linde_hybrid, kofman, kofman1, boyanovsky, greene, braden}. An alternative model postulates that the decay of the inflaton field occurs during the whole inflationary period, as in Warm Inflation (WI) scenarios ~(\cite{berera_fang, berera, berera_review, bastero_review, berera_spectral_index, taylor_berera, oliveira, oliveira1, delcampo, hall, bastero_gil, moss_xiong, moss, graham, delcampo2010, bastero1, moss1, setare2013, bastero2011, bartrum}, see also Refs.~\cite{fang1980, hosoya, moss_ewi, lonsdale, yokoyama, liddle_ewi}). Other models include multi-inflaton fields~\cite{garcia_bellido, randall, tsujikawa, parkinson, kadota, wands2007, bassett}, natural inflation~\cite{natural_inflation, natural_inflation1, natural_inflation2, mohanty, visinelli_NWI}, and chain inflation~\cite{freese2005, freese2005_1}; alternative mechanisms include the curvaton~\cite{enqvist2002, lyth2002, moroi2002, moroi2002_1}.

The only constraint on the reheating temperature of the Universe $\TRH \gtrsim 4{\rm ~MeV}$, valid in both the standard and WI models, comes from the requirement that a sufficiently large population of thermal neutrinos has to be produced~\cite{kawasaki, kawasaki1, hannestad, ichikawa}, since a modification of the neutrino number density would spoil the observed abundance of light elements, the observed CMBR and matter power spectra~\cite{ichikawa1, bernardis}, and the large-scale structure. 

As previously mentioned, inhomogeneities might have generated during inflation~\cite{bardeen1980, kodama, mukhanov1992}, either as quantum fluctuations of the inflaton (standard inflation scenarios~\cite{mukhanov, guth_pi, hawking, starobinsky1, bardeen1}), or as thermal fluctuations (WI scenarios~\cite{berera_fang, berera}). These cosmological fluctuations later evolve into perturbations of the matter and radiation power spectra that are observed in the CMBR. The evolution of the fields participating inflation is described by a set of coupled Boltzmann equations, whose first-order perturbation describes the evolution of inhomogeneities. To some extent, these fields can be treated as fluids, with a pressure related to the energy density by an equation of state. The formalism for interacting fluids in an expanding Friedmann-Robertson-Walker (FRW) metric has been extensively discussed in Refs.~\cite{kodama, mukhanov1992, malik, malik1}.

In this work, we present the equations describing the coupling of the inflaton to a relativistic field using the interacting fluids approach developed in Refs.~\cite{malik, malik1}. The equations for the bulk and the cosmological fluctuations are solved numerically and tested against the analytical approximation around the reheating period. The paper is organized as follows. In Sec.~\ref{Equations for the background fields}, we provide a numerical solution to the basic equations for the background fields, while in Sec.~\ref{Evolution of perturbations during inflation} we study the time evolution of the perturbations in the inflaton and radiation fields. Sec.~\ref{Stochastic perturbations} is devoted to the modifications induced in the equation by an additional stochastic term, which is usually considered in WI scenarios. Results are summarized in Sec.~\ref{Discussion and conclusions}.

\section{Equations for the background fields} \label{Equations for the background fields}

We consider a model in which the energy density of the inflaton field $\rho_\phi$ dominates the energy density of the Universe, decaying into relativistic matter with energy density $\rho_r$ at a rate $\Gamma$. We assume a FRW metric with cosmic time $t$ and scale factor $a = a(t)$,
\begin{equation} \label{line_element_0}
ds^2 = -dt^2 + a^2(t)\,\delta_{ij}\,dx^i\,dx^j.
\end{equation}
If radiation thermalizes on a time scale much shorter than $1/\Gamma$, the potential $U(\phi, T)$~\cite{bellini, matsuda}, that expresses the interaction between the inflaton and radiation, splits as $U(\phi,T) = U(\phi) + \rho_r$~\cite{berera, moss}. Denoting the zero-th order (background) term in the perturbation with a bar over a quantity, we write the background expressions for the energy density and pressure fields as
\begin{equation}
\begin{array}{l}
\displaystyle \bar{\rho}_\phi = \frac{1}{2}\,\dot{\bar{\phi}}^2 + U(\bar{\phi}),\quad\quad \bar{p}_\phi = \frac{1}{2}\,\dot{\bar{\phi}}^2 - U(\bar{\phi}),\\
\displaystyle \bar{\rho}_r = \frac{\pi^2}{30}\,g_*(T)\,T^4, \quad \quad \bar{p}_r = c_r^2\,\bar{\rho}_r,
\end{array}
\label{energy_pressure}
\end{equation}
where $c_r^2 = 1/3$ describes the equation of state for relativistic matter. Under these conditions, Friedmann equations read~\cite{weinberg_book}
\begin{equation}
\begin{array}{l}
\displaystyle H^2 =  \frac{8\pi\,G}{3}\,\bar{\rho} = \frac{8\pi\,G}{3}\left(\frac{1}{2}\dot{\bar{\phi}}^2+U+\bar{\rho}_r\right),\\
\displaystyle\\
\displaystyle \dot{H} = -4\pi\,G\,\left(\bar{p} + \bar{\rho}\right) = - 4\pi\,G\,\left(\dot{\bar{\phi}}^2 + \frac{4}{3}\,\bar{\rho}_r\right),
\end{array}
\label{friedmann}
\end{equation}
where $G$ is Newton's gravitational constant, $H = \dot{a}/a$ is the Hubble rate, and where a dot indicates a derivation with respect to the cosmic time $t$.

For each species $\alpha$, the continuity equation is~\cite{kodama}
\begin{equation}\label{cons_energy_alpha}
\begin{array}{l}
\displaystyle \dot{\bar{\rho}}_\alpha + 3H\,\left(\bar{p}_\alpha + \bar{\rho}_\alpha\right) = \bar{Q}_\alpha,\\
\end{array}
\end{equation}
where $\bar{Q}_\alpha$ is a source term describing the conversion between the species $\alpha$ and the other species accounted in the theory. To assure the conservation of total energy density in a co-moving volume, the sum of all source terms must satisfy
\begin{equation}
\sum_\alpha\,\bar{Q}_\alpha = 0.
\end{equation}
The conversion of the inflaton energy density into radiation energy density is thus described by the set of Boltzmann equations
\begin{equation}\label{cons_energy}
\begin{array}{l}
\displaystyle \dot{\bar{\rho}}_\phi + 3H\,\left(\bar{p}_\phi + \bar{\rho}_\phi\right) = -\bar{Q},\\
\displaystyle \dot{\bar{\rho}}_r + 3H\,\left(\bar{p}_r + \bar{\rho}_r\right) = \bar{Q}.
\end{array}
\end{equation}
At early times $t \ll 1/\Gamma$, the source term in the Boltzmann equation can be neglected, and the inflaton energy density is described by
\begin{equation} \label{no_source_phi}
\frac{d\bar{\rho}_\phi}{da} + \frac{3}{a}\,\left(\bar{p}_\phi + \bar{\rho}_\phi\right) = 0.
\end{equation}
For a pressure-less fluid $\bar{p}_\phi =0$, Eq.~(\ref{no_source_phi}) predicts $\bar{\rho}_\phi \propto a^{-3}$~\cite{erickcek}. In this paper instead, we use the expression $\bar{p}_\phi + \bar{\rho}_\phi = 2\epsilon\,\bar{\rho}_\phi/3$, where $\epsilon$ is the first slow-roll parameter defined in Eq.~(\ref{slow_roll_parameters}) below. With this expression, the energy density is constant during during $\phi$ domination as $\bar{\rho}_\phi \propto a^{-2\epsilon}$.

\begin{figure}[h!]
\centering
\includegraphics[width=15cm]{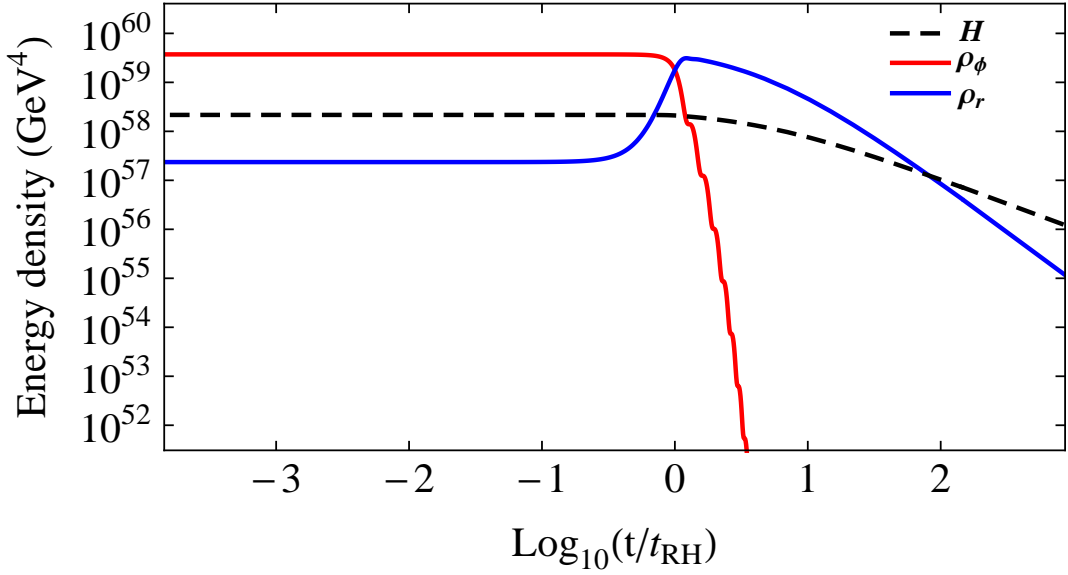}
\caption{The energy density for the inflaton field $\bar{\rho}_\phi$ (red line), the radiation field $\bar{\rho}_r$ (blue line), and the Hubble rate $\phi_0^2\,\Gamma H$, as a function of time $t$.}
\label{figure:background}
\end{figure}

Inserting the expressions for $\bar{\rho}_\phi$ and $\bar{p}_\phi$ given in Eq.~(\ref{energy_pressure}), together with the source term
\begin{equation} \label{source_term}
\bar{Q} = \Gamma\,\left(\bar{p}_\phi + \bar{\rho}_\phi\right),
\end{equation}
into Eq.~(\ref{cons_energy}), we obtain the evolution of the inflaton and radiation fields in a FRW metric,
\begin{equation} \label{eq_motion}
\begin{array}{l}
\displaystyle \ddot{\bar{\phi}} + \left(3H + \Gamma\right)\,\dot{\bar{\phi}} + U_\phi = 0,\\
\displaystyle \dot{\bar{\rho}}_r + 4H\,\bar{\rho}_r = \Gamma\,\dot{\bar{\phi}}^2,
\end{array}
\end{equation}
where $U_\phi = \partial U/\partial\bar{\phi}$. The second line in Eq.~(\ref{eq_motion}) corresponds to the Boltzmann equation for the massive scalar field $\phi$ decaying into a massless field $\chi$ via an interaction proportional to $\phi\,\chi^2$~\cite{braden}. Here, we do not account for a microscopical model for the radiation field, which is treated as a fluid. The set in Eq.~(\ref{eq_motion}) has been often presented various time in the literature. Examples include warm inflation~\cite{oliveira, oliveira1, hall, delcampo, graham, visinelli_NWI}, the decay of the inflaton~\cite{allahverdi, allahverdi1} or the curvaton field~\cite{gupta, matarrese}, or the early domination of a massive scalar field ~\cite{chung, erickcek}. We solve the set of Eq.~(\ref{eq_motion}) numerically, assuming that the inflaton potential be~\cite{lyth}
\begin{equation}\label{potential}
U(\phi) = \frac{\lambda}{4}\,\left(\phi^2 - \phi_0^2\right)^2,
\end{equation}
where $\lambda$ is a dimensionless parameter describing the strength of the potential, and $\phi_0$ is the value of the inflaton field at which the potential reaches its minimum value. Initial conditions for $\dot{\phi}$ and $\rho_r$ are obtained by first solving the set of Eq.~(\ref{eq_motion}) in the regime $\ddot{\phi} \ll H\,\dot{\phi}$ and $\dot{\rho}_r \ll H\,\rho_r$, starting from an initial configuration of the inflaton $\phi = \phi_i$. In Fig.~\ref{figure:background}, we show the energy densities $\bar{\rho}_\phi$ (red solid line) and $\bar{\rho}_r$ (blue solid line) as a function of time $t$, for the values of the parameters given in Table~\ref{table_parameters}. Also shown is the Hubble rate $H$, multiplied by the constant $\phi_0^2\,\Gamma$ in order to obtain the units of the energy density.

\begin{center}
\begin{table}[h!]
\centering
\begin{tabular}{l}
{\bf Quantity}\\
\hline\\
$\lambda=10^{-8}$\\
$\Gamma = 10^{13}{\rm ~GeV}$\\
$\phi_0 = 10^{17}{\rm ~GeV}$\\
$\phi_i = 0.01\,\phi_0$\\
\hline\\
\end{tabular}
\caption{Values for the parameters used in Fig.~\ref{figure:background}.}
\label{table_parameters}
\end{table}
\end{center}

\begin{figure}[h!]
\centering
\includegraphics[width=15cm]{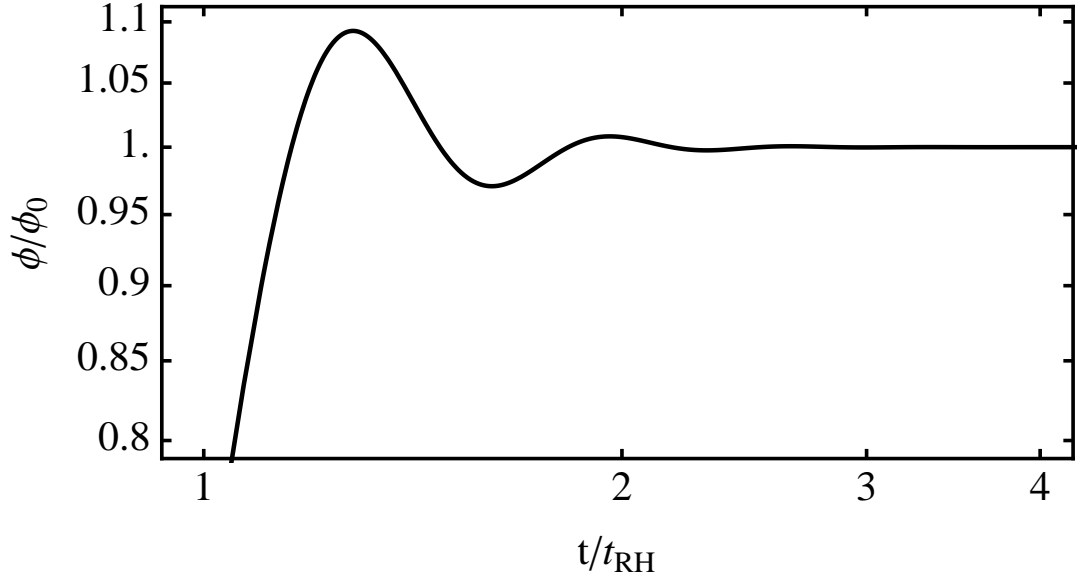}
\caption{The behavior of the inflaton field $\phi(t)$ around the time $\tRH$ at which the inflaton converts into relativistic degrees of freedom, as a function of time $t$.}
\label{figure:phi}
\end{figure}

Defining the reheating time $\tRH$ as the time at which
\begin{equation}
\bar{\rho}_\phi(\tRH) = \bar{\rho}_r(\tRH),
\end{equation}
the total energy density is dominated by the inflaton energy density for $t < \tRH$ , with a constant value $\bar{\rho}_\phi \approx U(\phi_i)$. When $t \approx \tRH$, the inflaton field starts oscillating around its equilibrium value $\phi_0$, so we write the expansion $\phi = \phi_0 + \varphi$, where $\varphi$ is a small perturbation. In this approximation, the inflaton potential is $U(\phi_0+\varphi) \approx \lambda\,\phi_0^2\,\varphi^2$, and Eq.~(\ref{eq_motion}) reads
\begin{equation} \label{eq_motion_tRH}
\begin{array}{l}
\displaystyle \ddot{\varphi} + \left[\sqrt{24\pi\,G}\,\left(\lambda\,\phi_0^2\,\varphi^2 + \frac{\dot{\varphi}^2}{2} + \bar{\rho}_r\right)^{1/2} + \Gamma\right]\,\dot{\varphi} + 2\lambda\,\phi_0^2\,\varphi = 0,\\
\displaystyle \dot{\bar{\rho}}_r + 4\,\sqrt{\frac{8\pi\,G}{3}}\,\left(\lambda\,\phi_0^2\,\varphi^2 + \frac{\dot{\varphi}^2}{2} + \bar{\rho}_r\right)^{1/2}\,\bar{\rho}_r -\Gamma\,\dot{\varphi}^2 = 0.
\end{array}
\end{equation}
Keeping only the leading terms in $\varphi$ and $\dot{\varphi}$, we obtain
\begin{equation} \label{eq_motion_tRH1}
\begin{array}{l}
\displaystyle \ddot{\varphi} + \Gamma\,\dot{\varphi} + 2\lambda\,\phi_0^2\,\varphi = 0,\\
\displaystyle \dot{\bar{\rho}}_r + 4\,\sqrt{\frac{8\pi\,G}{3}}\,\bar{\rho}_r^{3/2} = 0,
\end{array}
\end{equation}
with solution
\begin{equation} \label{eq_motion_tRHsol}
\begin{array}{l}
\displaystyle \varphi(t) \propto e^{-\Gamma\,t/2}\,\cos\left(\frac{t}{2}\,\sqrt{\Gamma^2 - 8\lambda\,\phi_0^2} \right),\\
\displaystyle \bar{\rho}_r(t) \propto t^{-2}.
\end{array}
\end{equation}
The inflaton field approaches its equilibrium value $\phi_0$ with an exponential decay; for the sake of illustration, in Fig.~\ref{figure:phi} we show a detail of the value of the inflaton field around $t = \tRH$. The dependence of the energy density of the relativistic field $\bar{\rho}_r(t) \propto t^{-2}$ is a standard result, which can also be inferred from the decoupled conservation equation,
\begin{equation} \label{eq_motion_rad}
\displaystyle \dot{\bar{\rho}}_r + 4H\,\bar{\rho}_r \approx 0,
\end{equation}
and the fact that $a(t) \propto t^{1/2}$ during the radiation epoch. From this result, it follows that the Hubble rate (the black dashed line in Fig.~\ref{figure:background}) is $H(t) = 1/2t$ when $t > \tRH$. 

\section{Evolution of perturbations during inflation} \label{Evolution of perturbations during inflation}

We consider the perturbations of the FRW background metric in Eq.~(\ref{line_element_0}), described by the metric in the longitudinal gauge \cite{bardeen1980, mukhanov1992}
\begin{equation} \label{line_element}
ds^2 = -\left(1 + 2\Psi\right)\,dt^2 + a^2(t)\,\delta_{ij}\,\left(1 - 2\Psi \right)\,dx^i\,dx^j,
\end{equation}
where $\Psi$ parametrizes the perturbations in the metric. Eq.~(\ref{line_element}) is valid whenever perturbations of the total energy-momentum tensor do not give rise to anisotropic stress. 

Following Eqs.~(2.15) and~(2.17) in Ref.~\cite{malik} with our notation and in the case in which the shear is zero, the expression for the perturbations in the energy density and pressure fields of a generic species $\alpha$ are
\begin{equation}
\delta \dot{\rho}_\alpha + 3H\,\left(\delta \rho_\alpha + \delta p_\alpha\right) - 3\left(\bar{\rho}_\alpha + \bar{p}_\alpha\right)\,\dot{\Psi} + a^{-2}\,\nabla^2\,\left(\bar{\rho}_\alpha + \bar{p}_\alpha\right)\,\delta u_\alpha - \bar{Q}_\alpha\,\Psi - \delta Q_\alpha = 0,\\
\label{perturbation_energy_malik}
\end{equation}
and
\begin{equation}
\delta p_\alpha + a^{-3}\,\left[a^3\,\left(\bar{\rho}_\alpha + \bar{p}_\alpha\right)\,\delta u_\alpha\right]^{\centerdot} + \left(\bar{\rho}_\alpha + \bar{p}_\alpha\right)\,\Psi = \bar{Q}_\alpha\,\delta u.
\label{perturbation_pressure_malik}
\end{equation}
Here, a dot indicates a derivation with respect to $t$, $\delta \rho_\alpha$ and $\delta p_\alpha$ are respectively the first-order perturbations in the energy density and pressure fields of the $\alpha$-fluid, $c^2_\alpha = \dot{p}_\alpha/\dot{\rho}_\alpha$, $\delta u$ is the total covariant velocity perturbation, and $\nabla^2$ is the co-moving spatial Laplacian. Eq.~\eqref{perturbation_pressure_malik} can be alternatively reformulated as
\begin{equation}
\delta \dot{u}_\alpha + \left[\frac{\bar{Q}_\alpha}{\bar{\rho}_\alpha + \bar{p}_\alpha}\,\left(1 + c^2_\alpha\right)-3\,c^2_\alpha\,H\right]\,\delta u_\alpha + \Psi + \frac{1}{\bar{\rho}_\alpha + \bar{p}_\alpha}\,\left[\delta p_\alpha -\bar{Q}_\alpha\,\delta u\,\right] = 0.
\label{perturbation_pressure_malik_old}
\end{equation}
Defining the Fourier transform of a generic quantity $F({\bf x}, t)$ as
\begin{equation}
\begin{array}{l}
\displaystyle F_q \equiv F({\bf q}, t) = \frac{1}{(2\pi)^3}\,\int\,d^3x\,e^{-i\,{\bf q}\cdot{\bf x}}\,F({\bf x}, t),
\end{array}
\end{equation}
we take the Fourier transform of the equations for density and pressure perturbations, and set $\nabla^2 \to -q^2$. In momentum space, Eq.~(\ref{perturbation_energy_malik}) for the inflaton and the radiation fields reads
\begin{equation}
\begin{split}
\begin{array}{l}
\displaystyle \delta \dot{\rho}_{\phi q} + 3H\,(\delta \rho_{\phi q} + \delta p_{\phi q}) + \left(\bar{\rho}_\phi+\bar{p}_\phi\right)\,\left[\Gamma\,\Psi_q - 3\dot{\Psi}_q - \frac{q^2}{a^2}\delta u_{\phi q}\right] = -\delta Q,\\
\displaystyle \delta \dot{\rho}_{r q} +3H(\drq + \dpq) - \bar{\rho}_r\,\left[\frac{4}{3}\frac{q^2}{a^2}\,\duq + 4\dot{\Psi}_q\right] - \Gamma\,\left(\bar{p}_\phi + \bar{\rho}_\phi\right)\, \Psi_q = \delta Q.
\end{array}
\label{perturbation_energy1}
\end{split}
\end{equation}
The first-order perturbation of the source terms in Eq.~(\ref{source_term}) is
\begin{equation}\label{cold_perturbations}
\delta Q = \Gamma\,\left(\delta p_\phi + \delta \rho_\phi\right),
\end{equation}
while the term $\bar{Q}\,\Psi_q$ containing the perturbations in the gravitational field has already been included in Eq.~(\ref{perturbation_energy1}) which, once the term $\delta Q$ in Eq.~(\ref{cold_perturbations}) has been substituted, reads
\begin{equation}
\begin{split}
\begin{array}{l}
\displaystyle \delta \dot{\rho}_{\phi q} + \left(3H+\Gamma\right)\,\left(\delta \rho_{\phi q} + \delta p_{\phi q}\right) + \left(\bar{\rho}_\phi+\bar{p}_\phi\right)\,\left[\Gamma\,\Psi_q - 3\dot{\Psi}_q - \frac{q^2}{a^2}\delta u_{\phi q}\right] = 0,\\
\displaystyle \delta \dot{\rho}_{r q} +3H\,(\drq+\dpq) - \bar{\rho}_r\,\left[\frac{4}{3}\frac{q^2}{a^2}\,\duq + 4\dot{\Psi}_q\right] - \Gamma\,\left(\bar{p}_\phi + \bar{\rho}_\phi\right)\, \Psi_q = \Gamma\,\left(\delta \rho_{\phi q} + \delta p_{\phi q}\right).
\end{array}
\label{perturbation_energy2}
\end{split}
\end{equation}
For velocity perturbations, we rewrite Eq.~(\ref{perturbation_pressure_malik_old}) for the inflaton and radiation fields as
\begin{equation}\label{perturbation_pressure1}
\begin{array}{l}
\displaystyle \delta \dot{u}_{\phi q} - \left(3H + \Gamma\right)\,c^2_\phi\,\delta u_{\phi q} + \Psi_q + \Gamma\,\left(\delta u_q - \delta u_{\phi q}\right) + \frac{\delta p_{\phi q}}{\bar{\rho}_\phi + \bar{p}_\phi} = 0,\\
\displaystyle\\
\displaystyle \frac{4}{3}\,\bar{\rho}_r\,\left(\delta \dot{u}_{r q} - H\,\delta u_{r q} + \Psi_q \right) + \Gamma\,\left(\bar{\rho}_\phi + \bar{p}_\phi\right)\,\left(\frac{4}{3}\,\delta u_{r q} - \delta u_q\right) + \delta p_{r q} = 0.\\
\end{array}
\end{equation}
Perturbations in the gravitational field, described by the field $\Psi$ appearing in the metric in Eq.~\eqref{line_element}, are described by~\cite{malik}
\begin{equation}
\dot{\Psi}_q + H\,\Psi_q + 4\pi\,G \,\left[\frac{4}{3}\,\bar{\rho}_r\,\delta u_{r q} + \left(\bar{\rho}_\phi + \bar{p}_\phi\right)\, \delta u_{\phi q}\right] = 0.
\label{perturbation_field1}
\end{equation}
Following Weinberg~\cite{weinberg_book}, the perturbations in the energy density, pressure, and velocity of the inflation field are related to the perturbations in the scalar field $\delta \phi$ by
\begin{equation}
\begin{array}{l}
\displaystyle \delta \rho_{\phi q} = \dot{\bar{\phi}}\,\delta\dot{\phi}_q + U_\phi\,\delta\phi_q - \dot{\bar{\phi}}^2\,\Psi_q,\\
\displaystyle \delta p_{\phi q} = \dot{\bar{\phi}}\,\delta\dot{\phi}_q - U_\phi\,\delta\phi_q - \dot{\bar{\phi}}^2\,\Psi_q,\\
\displaystyle \delta u_{\phi q} = -\frac{\delta \phi_q}{\dot{\bar{\phi}}}.
\end{array}
\label{field_perturbations}
\end{equation}
In addition, we assume that perturbations in the radiation fluid are adiabatic, setting $\dpq = c_r^2\,\drq = \drq/3$. Combining the set in Eq.~(\ref{field_perturbations}) with the expressions for the energy density and pressure fields in Eqs.~(\ref{perturbation_energy2}),~(\ref{perturbation_pressure1}) and~(\ref{perturbation_field1}), we obtain
\begin{equation} \label{perturbation_eq_phi}
\delta \ddot{\phi}_q + \left(3H +\Gamma\right)\,\ddfq + \left(U_{\phi\phi} + \frac{q^2}{a^2}\right)\,\dfq = 4\,\dot{\bar{\phi}}\,\dot{\Psi}_q - \left(2\,U_\phi + \Gamma\,\dot{\bar{\phi}}\right)\,\Psi_q,
\end{equation}
\begin{equation} \label{perturbation_eq_r}
\delta \dot{\rho}_{r q} +4H\,\drq - \frac{4}{3}\frac{q^2}{a^2}\,\bar{\rho}_r\,\delta u_{r q} = 4\bar{\rho}_r\,\dot{\Psi}_q + \Gamma\,\dot{\bar{\phi}}\,\left(2\delta\dot{\phi}_q - \dot{\bar{\phi}}\,\Psi_q\right),
\end{equation}
\begin{equation} \label{perturbation_eq_u}
\frac{4}{3}\,\bar{\rho}_r\,\left(\delta \dot{u}_{r q} - H\,\delta u_{r q} + \Psi_q \right) + \Gamma\,\dot{\bar{\phi}}\,\left(\frac{4}{3}\,\dot{\bar{\phi}}\,\delta u_{r q} + \delta \phi_q\right) + \frac{1}{3}\,\delta \rho_{r q} = 0,
\end{equation}
\begin{equation} \label{perturbation_eq_psi}
\dot{\Psi}_q + H\,\Psi_q + 4\pi\,G \,\left(\frac{4}{3} \bar{\rho}_r\,\delta u_{r q} - \dot{\bar{\phi}}\,\delta \phi_q\right) = 0.
\end{equation}
In the literature, velocity perturbations in the radiation fluid have often been expressed in terms of the scalar potential velocity $v_r$, related to the covariant velocity perturbation used here by $\delta u_{r q} = a\,v_r/q$. When substituting for $v_r$, Eq.~(\ref{perturbation_eq_u}) reads
\begin{equation} \label{scalar_potential_r}
\dot{v}_r + \frac{\Gamma\,\dot{\bar{\phi}}^2}{\bar{\rho}_r}\,v_r +\frac{q}{a}\left(\Psi_q + \frac{3\Gamma\,\dot{\bar{\phi}}}{4\bar{\rho}_r}\,\delta \phi_q + \frac{\delta \rho_{r q}}{4\bar{\rho}_r}\right) = 0.
\end{equation}
In their Eq.~(A20), Moss and Xiong~\cite{moss_xiong} quote the same expression as our Eq.~(\ref{scalar_potential_r}), once set their $\alpha = \Psi_q$ and their $J = -\Gamma\,\dot{\bar{\phi}}\,\delta \phi_q$ as in their Eq.~(A13). Oliveira and Joras~\cite{oliveira} quote the same as our Eq.~(\ref{scalar_potential_r}) in the third line of their Eq.~(A2); the second line of their Eq.~(A2) agrees with our Eq.~(\ref{perturbation_eq_phi}) once set $\Gamma'=0$, while their first line in Eq.~(A2) shows a term with an extra factor of three with respect to our Eq.~(\ref{perturbation_eq_r}). Our Eq.~(\ref{perturbation_eq_phi}) shows all terms on the RHS of the equation with the opposite sign with respect to Eq.~(61) quoted in Hall {\it et al.}~\cite{hall}.

\begin{figure}[h!]
\centering
\includegraphics[width=12cm]{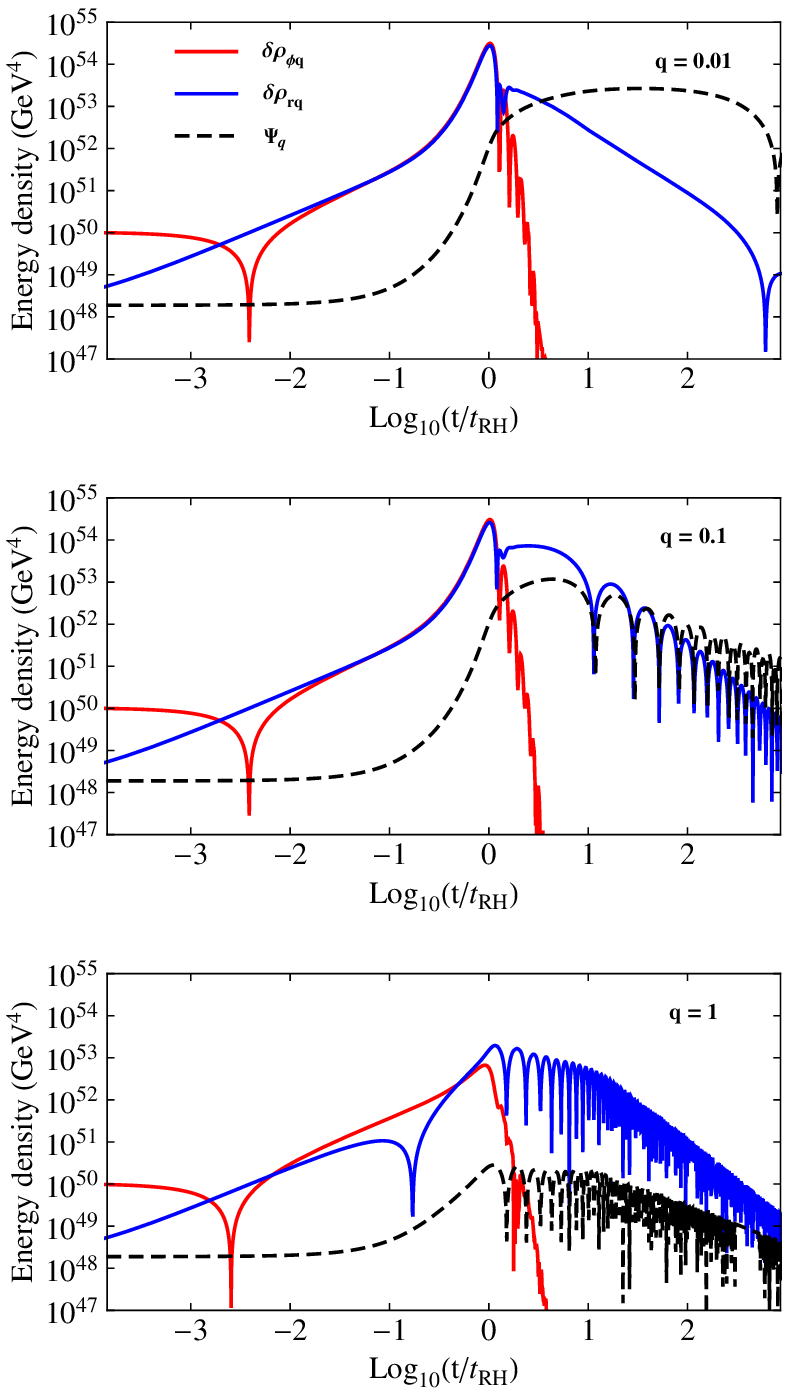}
\caption{Perturbation in the energy density of the (red solid line) and radiation (blue solid line) and gravitational perturbations (black dashed line) in order of $\Gamma^2\phi_0^2$ obtained from the numerical resolution of the set of Eqs.~\eqref{perturbation_eq_phi},~\eqref{perturbation_eq_r},~\eqref{perturbation_eq_u},~\eqref{perturbation_eq_psi}, with the parameters in Table~\ref{table_parameters} and for $q = 0.01$ (top figure), $q=0.1$(middle figure), and $q=1$ (bottom figure).}
\label{figure:pertQ}
\end{figure}

To obtain the initial conditions for $\delta \phi_q$ and $\Psi_q$, we perform a WKB expansion using the fact that, at early times, the term $q/a(t)$ is much larger than $H$, and radiation can be neglected. Introducing the conformal time $d\tau = dt/a(t)$, we write
\begin{equation}
\displaystyle \delta \phi_q = f\,\exp\left(-i\,q\,\tau\right),\quad\hbox{and}\quad \Psi_q = g\,\exp\left(-i\,q\,\tau\right),
\end{equation}
where $f(t)$ and $g(t)$ are slowly-varying functions. Substituting these expressions in Eqs.~(\ref{perturbation_eq_phi}) and~(\ref{perturbation_eq_psi}), and keeping the leading terms in $q/a$ only, we obtain
\begin{equation} \label{perturbation_eq1}
\displaystyle f \sim a^{-3/2}, \quad \hbox{and} \quad g = \frac{4\pi\,i\,G\,\dot{\bar{\phi}}\,a}{q}\,f.
\end{equation}
Initial conditions for $\delta \rho_{r q}$ and $\delta u_{r q}$ are obtained by solving Eqs.~\eqref{perturbation_eq_phi} and~\eqref{perturbation_eq_psi} for these variables, neglecting their time dependence and given the initial conditions for $\delta \phi_q$ and $\Psi_q$. We solve the set of Eqs.~\eqref{perturbation_eq_phi}~\eqref{perturbation_eq_r}~\eqref{perturbation_eq_u}~\eqref{perturbation_eq_psi} numerically using these initial conditions and the numerical solution to the background Eq.~(\ref{eq_motion}). The scale factor $a(t)$ that appears in the set of equations is obtained as the numerical solution of the differential equation $\dot{a} = H\,a$, where the Hubble rate $H$ is defined by the Friedmann Eq.~(\ref{friedmann}). In Fig.~\ref{figure:pertQ}, we show results for the numerical resolution of this set of equations with the values for the parameters as in Table~\ref{table_parameters}, for the values $q = 0.01$ (a), $q = 0.1$ (b), $q = 1$ (c).

After the reheating stage, the radiation perturbations dominate over the inflaton perturbations. The first-order perturbations in the radiation field oscillate with a frequency that increases with $q$. To explain this oscillatory behavior, we set $\Gamma = 0$, and we introduce the variable $\delta_{r q} = \delta \rho_{r q}/\bar{\rho}_r$, so that Eqs.~\eqref{perturbation_eq_r},~\eqref{perturbation_eq_psi}, and~\eqref{scalar_potential_r}, when $t > \tRH$ are simplified as
\begin{equation} \label{perturbation_eq_1}
\begin{array}{l}
\displaystyle \dot{\delta}_{r q} - \frac{4}{3}\frac{q}{a}\,v_{r q} - 4\dot{\Psi}_q = 0,\\
\displaystyle \dot{v}_r+\frac{q}{a}\left(\Psi_q + \frac{1}{4}\,\delta_{r q}\right) = 0,\\
\displaystyle \dot{\Psi}_q + H\,\Psi_q + 2H^2\,\frac{a}{q}\,v_{r q} = 0.
\end{array}
\end{equation}
Neglecting the perturbations in the gravitational potential $\Psi_q$, we find that both $\delta_{rq}$ and $v$ follow the same differential equation,
\begin{equation} \label{perturbation_eq_X}
\displaystyle \ddot{x} + H\,\dot{x} + \frac{q^2}{3a^2}\,x = 0,
\end{equation}
where $x(t)$ expresses either $\delta_{rq}$ or $v$. For a radiation-dominated cosmology, setting
\begin{equation} \label{perturbation_eq_solutionX}
H = \frac{1}{2t}, \quad a = \aRH\,\sqrt{\frac{t}{\tRH}},\quad Q^2 = \frac{q^2}{3\,\aRH^2},
\end{equation}
this differential equation has solution
\begin{equation}
x(t) = x_1\,\sin \left(2 Q\sqrt{\frac{t}{\tRH}}\right) + x_2\,\cos \left(2 Q\sqrt{\frac{t}{\tRH}}\right),
\end{equation}
where $x_1$ and $x_2$ are arbitrary constants.

\begin{figure}[h!]
\centering
\includegraphics[width=12cm]{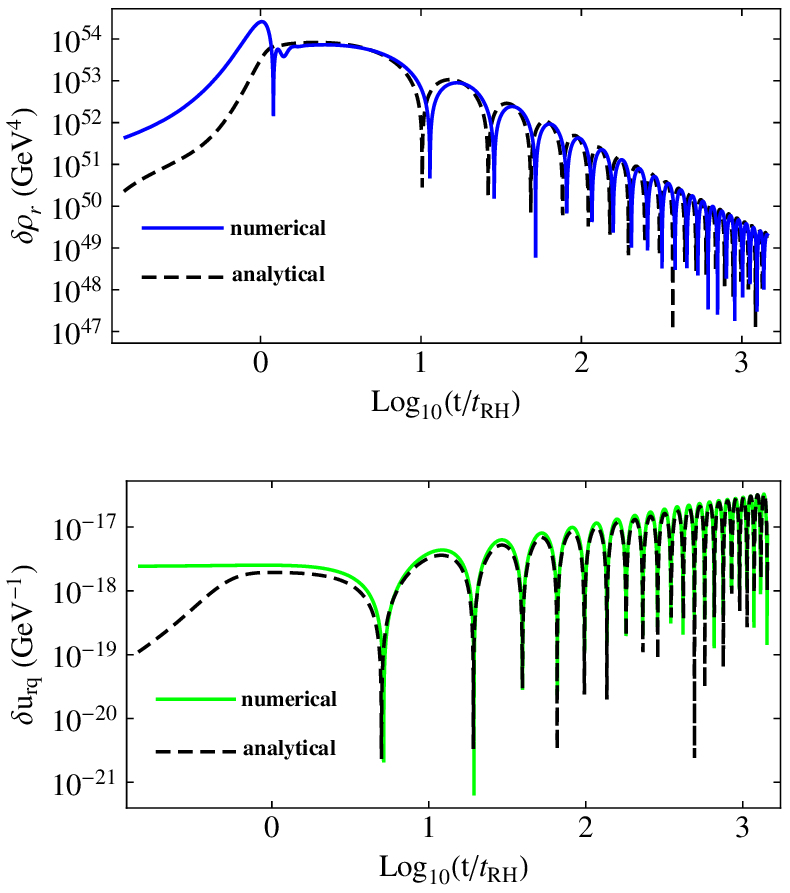}
\caption{Top: Density perturbations in the radiation fluid $\delta \rho_{r q}$ obtained with the numerical resolution of Eq.~\eqref{perturbation_eq_r} (blue solid line) and with the analytical solution of Eq.~\eqref{perturbation_eq_X} with $x(t) = \delta_{rq}$ (black dashed line). Bottom: Velocity perturbations in the radiation fluid $\delta u_{r q}$ obtained with the numerical resolution of Eq.~\eqref{perturbation_eq_u} (green solid line) and with the analytical solution of Eq.~\eqref{perturbation_eq_X} with $x(t) = v$ (black dashed line).}
\label{figure:pertvR}
\end{figure}

In Fig.~\ref{figure:pertvR}, we compare the solution to the expression in Eq.~\eqref{perturbation_eq_solutionX} in a radiation-dominated Universe (black dashed line) for the case of the density perturbations (top) and velocity perturbations (bottom), with the full numerical solution obtained from Eqs.~\eqref{perturbation_eq_r} (blue solid line) and~\eqref{perturbation_eq_u} (green solid line), setting $q = 0.1$. Once the matching initial conditions have been chosen, the analytical solution captures the oscillatory behavior and the trend obtained with the numerical resolution.

\section{Effects of a viscosity term} \label{dissipative_terms}

\begin{figure}[h!]
\centering
\includegraphics[width=15cm]{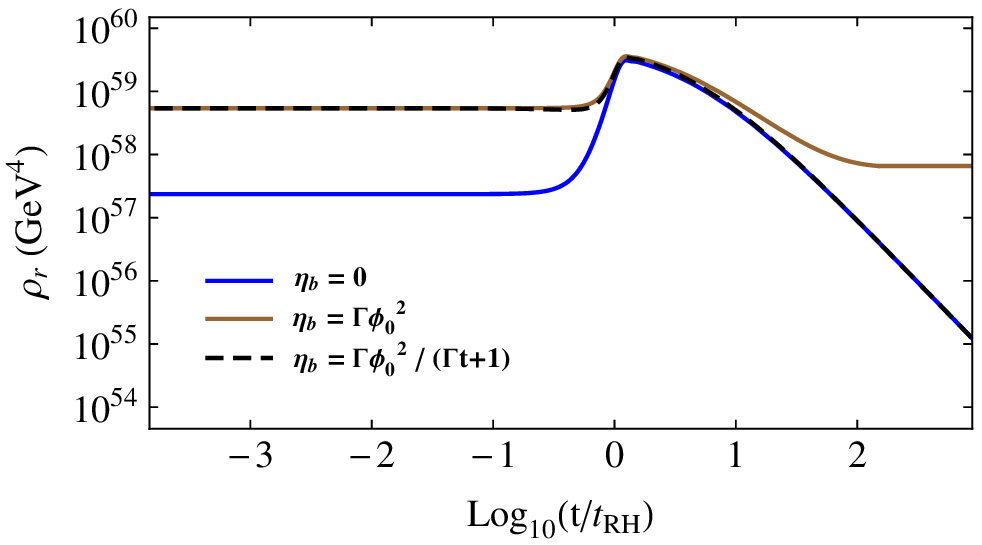}
\caption{The energy density in the radiation field $\bar{\rho}_r$, obtained from solving the set of Eqs.~\eqref{eq_motion1} with three different expressions for the viscosity term. For the viscosity coefficient, we set $\eta_b~=~0$ (blue solid line), $\eta_b~=~\Gamma\,\phi_0^2$ (brown solid line), and $\eta_b~=~\Gamma\phi_0^2/(\Gamma t+1)$ (black dashed line).}
\label{figure:backgroundviscosity}
\end{figure}

In the non-relativistic theory, the Navier-Stokes equation describes the evolution of the velocity field $\vv$ for a fluid of pressure $p$ and density $\rho$. In formulas~\cite{padmanabhan_book},
\begin{equation} \label{navier_stokes}
\frac{d\vv}{dt} = -\frac{1}{\rho}\,\n\,p - \n\cdot \PP,
\end{equation}
where the stress tensor $\PP$ introduces a dissipative behavior in the fluid motion, and satisfies
\begin{equation} \label{dissipative}
\n\cdot\PP = -\eta_s\,\n^2\,\vv - \left(\eta_b + \frac{\eta_s}{3}\right)\,\n\,\left(\n\cdot\vv\right).
\end{equation}
In the last expression, the divergence of $\vv$ can be written in terms of the identity
$$\n\,\left(\n\cdot\vv\right) = \n^2\,\vv + \n\times\n\times\vv.$$
Assuming that the vorticity of the fluid is zero, $\n\times\vv = 0$, which is a valid assumption for first-order cosmological perturbations~\cite{kodama}, allows us to write the velocity field as the gradient of a scalar field as $\vv = \n\,u$, where the scalar velocity potential $u$ is the non-relativistic analogue of the velocity potential $\delta u_\alpha$ introduced in Eq.~\eqref{perturbation_pressure_malik}. With this substitution, the Navier-Stokes Eq.~\eqref{navier_stokes} reads
\begin{equation}
\n\,\frac{du}{dt} = -\frac{1}{\rho}\,\n \,p + \eta_0\,\n\left(\n^2\, u\right),\quad\hbox{with}\quad \eta_0 = \frac{4}{3}\,\eta_s + \eta_b.
\end{equation}
So far the reviewing. We now discuss how viscosity enter the equation for a relativistic field. In Eq.~\eqref{eq_motion}, viscosity acts as an effective pressure, once we substitute~\cite{delcampo, bastero2011, bastero12a}
\begin{equation} \label{substitute_p}
\bar{p}_r \to \bar{p}_r  + \Pi_0,
\end{equation}
where $\Pi_0 = -3H\,\eta_b$. This modification alters the set of coupled equations for the background as
\begin{equation} \label{eq_motion1}
\begin{array}{l}
\displaystyle \ddot{\bar{\phi}} + \left(3H + \Gamma\right)\,\dot{\bar{\phi}} + U_\phi = 0,\\
\displaystyle \dot{\bar{\rho}}_r + 3H\,\left(\frac{4}{3}\,\bar{\rho}_r + \Pi_0\right) = \Gamma\,\dot{\bar{\phi}}^2.
\end{array}
\end{equation}
We find that the inclusion of the viscous term in Eq.~\eqref{eq_motion1} drastically affects the behavior of the solution for $\bar{\rho}_r$ when the scalar field has decayed at $t \gg 1/\Gamma$. In fact, in this regime the second line of Eq.~\eqref{eq_motion1} yields to a constant value of the radiation energy density,
\begin{equation} \label{viscosity_radiation_rho}
\bar{\rho}_r = \frac{27\,\pi\,G}{2}\,\eta_b^2,
\end{equation}
where we have used the fact that for a radiation-dominated cosmology the Hubble rate is $H^2 = 8\pi\,G\,\bar{\rho}_r/3$. A possible solution to this problem is found by considering a time-dependent viscous term, for example the usual result $\bar{\rho}_r \sim 1/t^2$ when $t \gg 1/\Gamma$ is recovered by setting $\eta_b \sim 1/t$. These results are confirmed in Fig.~\ref{figure:backgroundviscosity}, where we show the numerical resolution of Eq.~\eqref{eq_motion1}, with the values in Table~\ref{table_parameters} and with the choices for the viscosity coefficient as $\eta_b = 0$ (blue solid line), $\eta_b = \Gamma\,\phi_0^2$ (brown solid line), and $\eta_b = \Gamma\,\phi_0^2/(\Gamma\,t+1)$ (black dashed line). Notice that the blue solid line for $\bar{\rho}_r$ in Fig.~\ref{figure:backgroundviscosity} is the same as that obtained previously and shown in Fig.~\ref{figure:background}. Since $\eta_b$ enters the equation of motion as a negative pressure, its effect is that of enhancing the energy density of the radiation field. The viscosity term affects initial conditions, since they are obtained from Eq.~\eqref{eq_motion1} once $\ddot{\phi}$ and $\dot{\bar{\rho}}_r$ have been neglected. In particular, the solution with $\eta_b = \Gamma\,\phi_0^2(\Gamma t+1)^{-1}$ behaves like the one with $\eta_b~=~\Gamma\,\phi_0^2$ for $t \ll 1/\Gamma$, and like the one with $\eta_b = 0$ for $t \gg 1/\Gamma$.

\begin{figure}[h!]
\centering
\includegraphics[width=12cm]{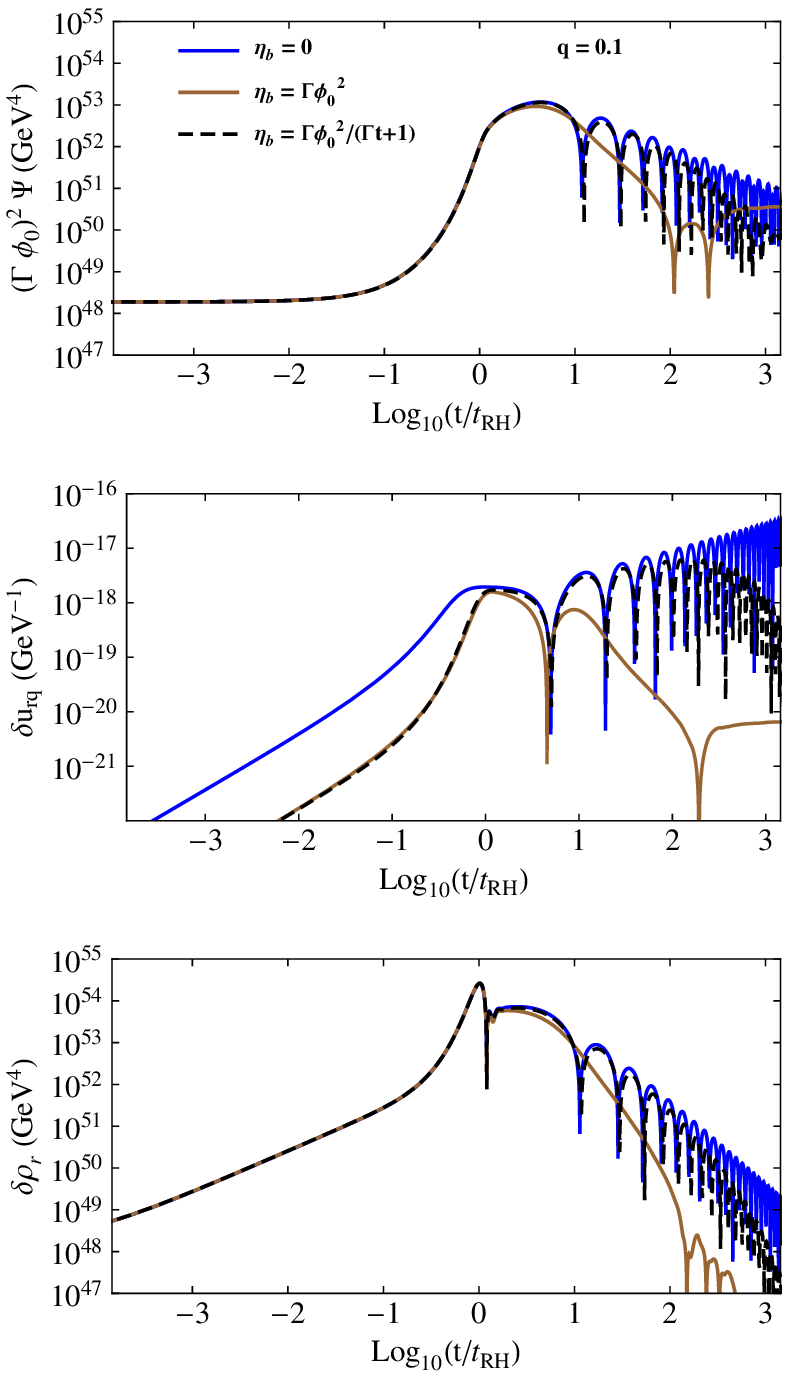}
\caption{Perturbations in the gravitational field $\Psi_{q}$ (top figure), in the velocity perturbation $\delta u_{rq}$ (middle figure) and in the perturbations of the radiation energy density $\delta \rho_{rq}$ (bottom figure), obtained by solving the set of Eqs.~\eqref{perturbation_eq_phi_viscosity1}-\eqref{perturbation_field_viscosity1}. For the viscosity coefficient, we set $\eta_b~=~0$ (blue solid line), $\eta_b~=~\Gamma\,\phi_0^2$ (brown solid line), and $\eta_b~=~\Gamma\phi_0^2/(\Gamma t+1)$ (black dashed line).}
\label{figure:eta}
\end{figure}

The effects of the dissipative terms into the relativistic expression for momentum and energy density perturbations have been extensively discussed in the literature~\cite{delcampo, delcampo2010, bastero2011, bastero1, setare2013, bastero12a}. Viscosities in the radiation fluid are included in Eq.~\eqref{perturbation_energy_malik} by substituting the pressure as in Eq.~\eqref{substitute_p}, while pressure perturbations modify as
\begin{equation} \label{delta_Pi}
\delta p_{rq} \to \delta p_{rq} + \delta \Pi_0.
\end{equation}
This gives 
\begin{equation}
\delta \dot{\rho}_{r q} + 3H\,(\drq+\dpq + \delta \Pi_0 ) - \left(\bar{\rho}_r+\bar{p}_r+\Pi_0\right)\,\left(\frac{q^2}{a^2}\,\duq + 3\dot{\Psi}_q\right) = \Gamma\,\dot{\bar{\phi}}\,\left(2\delta\dot{\phi}_q - \dot{\bar{\phi}}\,\Psi_q\right).
\label{perturbation_energy_viscosity}
\end{equation}
In order to derive the momentum equation for $\delta u_{rq}$, we add to the matter-energy tensor $T_{\mu\nu}$ of the system the stress tensor $\Pi_{\mu\nu}$, whose non-zero components are~\cite{bastero1}
\begin{equation} \label{def_viscosity}
\Pi_{ij}^{(\alpha)} = -\eta_0\,\nabla_i\,\nabla_j \,\delta u_\alpha,
\end{equation}
so that $\nabla_i \,\Pi_{ij}^{(\alpha)} = -\eta_0\,\nabla_j \,\nabla^2 \delta u_\alpha$. Eq.~\eqref{def_viscosity} can also be derived from the definition in Eq.~\eqref{dissipative} with the assumptions of zero vorticity. With the substitutions for the pressure term in Eq.~\eqref{substitute_p} and its perturbation in Eq.~\eqref{delta_Pi}, the momentum Eq.~\eqref{perturbation_pressure_malik} reads
\begin{equation}
\dpq + \delta \Pi_0 + a^{-3}\,\left[a^3\,\left(\bar{\rho}_r + \bar{p}_r + \Pi_0\right)\,\delta u_{rq}\right]^{\centerdot} + \left(\bar{\rho}_r + \bar{p}_r + \Pi_0\right)\,\Psi_q + \eta_0\,\frac{q^2}{a^2}\,\delta u_{rq}= \bar{Q}_r\,\delta u_q.
\label{perturbation_pressure_viscosity}
\end{equation}
Finally, Eq.~\eqref{perturbation_field1} giving the perturbations in the gravitational field modifies as
\begin{equation}
\dot{\Psi}_q + H\,\Psi_q + 4\pi\,G \,\left[\left(\frac{4}{3}\,\bar{\rho}_r + \Pi_0\right)\,\delta u_{r q} - \dot{\phi}\,\delta \phi_q\right] = 0.
\label{perturbation_field_viscosity}
\end{equation}
Summing up, Eqs.~\eqref{perturbation_eq_phi}-\eqref{perturbation_eq_psi} in the case where the viscosity is included as $\Pi_0 = -3H\,\eta_b$, with perturbations $\delta \Pi_0 = -3\left(H\,\Psi_q + \dot{\Psi}_q\right)\,\eta_b$, read
\begin{equation} \label{perturbation_eq_phi_viscosity1}
\delta \ddot{\phi}_q + \left(3H +\Gamma\right)\,\ddfq + \left(U_{\phi\phi} + \frac{q^2}{a^2}\right)\,\dfq = 4\,\dot{\bar{\phi}}\,\dot{\Psi}_q - \left(2\,U_\phi + \Gamma\,\dot{\bar{\phi}}\right)\,\Psi_q,
\end{equation}
\begin{equation}
\delta \dot{\rho}_{r q} + 4H\,\drq -9H^2\,\eta_b\,\Psi_q - \left(\frac{4}{3}\bar{\rho}_r - 3H\,\eta_b\right)\,\frac{q^2}{a^2}\,\duq = 4\,\bar{\rho}_r\,\dot{\Psi}_q + \Gamma\,\dot{\bar{\phi}}\,\left(2\delta\dot{\phi}_q - \dot{\bar{\phi}}\,\Psi_q\right),
\label{perturbation_energy_viscosity1}
\end{equation}
\begin{equation}
\frac{1}{3}\drq - 3\eta_b\,\left(2H\,\Psi_q + \dot{\Psi}_q\right) + \frac{4}{3}\,\bar{\rho}_r\,\Psi_q + a^{-3}\,\left[a^3\,\left(\frac{4}{3}\,\bar{\rho}_r -3H\eta_b\right)\,\delta u_{rq}\right]^{\centerdot} + \eta_0\,\frac{q^2}{a^2}\,\delta u_{rq} + \Gamma\,\dot{\bar{\phi}}\,\delta\phi_q = 0,
\label{perturbation_pressure_viscosity1}
\end{equation}
\begin{equation}
\dot{\Psi}_q + H\,\Psi_q + 4\pi\,G \,\left[\left(\frac{4}{3}\,\bar{\rho}_r -3H\,\eta_b \right)\,\delta u_{r q} - \dot{\phi}\,\delta \phi_q\right] = 0.
\label{perturbation_field_viscosity1}
\end{equation}
For completeness, we have included Eq.~\eqref{perturbation_eq_phi_viscosity1} although it has not been modified from Eq.~\eqref{perturbation_eq_phi} by the inclusion of viscosities. This set of equations describes the evolution of perturbations in the presence of viscous terms $\eta_b$ and $\eta_0$, and reduces to the results in the previous Section when $\eta_b = \eta_0 = 0$. We solve numerically the set of Eqs.~\eqref{perturbation_eq_phi_viscosity1}-\eqref{perturbation_field_viscosity1}, with the background for the inflaton field and $\bar{\rho}_r$ given by the solution to Eq.~\eqref{eq_motion1}. Results are shown in Fig.~\ref{figure:eta}, where we fixed $q = 0.1$. For the viscosity coefficients we set $\eta_0 = \eta_b$ and we choose $\eta_b = 0$ (blue line), corresponding to the solution described in Sec.~\ref{Evolution of perturbations during inflation}, $\eta_b = \Gamma\,\phi_0^2$ (brown line), which for the background gives the results in Eq.~\eqref{viscosity_radiation_rho} for large values of $t$, and $\eta_b = \Gamma\phi_0^2/(\Gamma t+1)$ (black dashed line), which allows us to retrieve the $\bar{\rho}_r \sim 1/t^2$ for large values of $t$. The presence of a constant viscosity term damps oscillations in all three fields shown.

Recently, Bastero-Gil~{\it et al.}~\cite{bastero1} proposed a model which uses the formalism of Hwang and Noh~\cite{hwang2002} for cosmological perturbations and Landau theory for the statistical fluctuations in the radiation fluid~\cite{landau}. Contrary to our assumption in Eq.~\eqref{energy_pressure}, where we indirectly assumed that the entropy density $s$ of the radiation fluid is a function of temperature only, in Ref.~\cite{bastero1} the entropy density is a function of both $T$ and $\phi$. The work in Ref.~\cite{bastero1} includes additional terms in the expression for the perturbations of the radiation energy density and velocity such as the viscosity coefficients introduced before $\eta_s$ and $\eta_b$, a possible perturbation in the dissipation coefficient $\delta\Gamma_q$, and stochastic sources ${\bf P}_q$, $\xi^{(r)}_q$, and $\xi^{(\phi)}_q$ (we discuss stochastic perturbations in Sec.~\ref{Stochastic perturbations} below). We have checked that imposing $\delta\Gamma_q = \bP = \xi^{(r)}_q = \xi^{(\phi)}_q = s_{,\phi} = 0$ in Eqs. (3.20), (3.24), (3.27), (3.28), and (3.30) of Ref.~\cite{bastero1} allows us to recover our set of Eqs.~\eqref{perturbation_energy_viscosity1}-\eqref{perturbation_field_viscosity1}.

\section{Slow-roll regime with a stochastic source}\label{Stochastic perturbations}

Inflation occurs if the potential $U(\phi)$ is sufficiently flat and much larger than all other forms of energy, so that the Hubble expansion rate $H$ is constant. During this period, which is known as the slow-roll regime of the inflaton field, higher derivatives in Eq.~(\ref{eq_motion}) can be neglected,
\begin{equation}\label{slow_roll_conditions}
\ddot{\phi} \ll H \,\dot{\phi},\quad\hbox{and}\quad \dot{\rho}_r \ll H\,\rho_r.
\end{equation}
In this regime, the Friedmann Eq.~(\ref{friedmann}) and Eq.~(\ref{eq_motion}) for the motion of the inflaton field read
\begin{equation}
\begin{array}{l}
\displaystyle H^2 \simeq \frac{8\pi G}{3}\,U,\\
\displaystyle \\
\displaystyle \dot{\phi} \simeq -\frac{U_\phi}{3H+\Gamma},\\
\displaystyle \\
\displaystyle \rho_r \simeq \frac{\Gamma\,\dot{\phi}^2}{4H},
\end{array}
\label{eq_motion_slow_roll}
\end{equation}
where we use the symbol ``$\simeq$'' for an equality that holds only in the slow-roll regime. Defining the slow-roll parameters,
\begin{equation}\label{slow_roll_parameters}
\epsilon = \frac{1}{16\pi G}\left(\frac{U_\phi}{U}\right)^2, \quad \eta = \frac{1}{8\pi G}\,\frac{U_{\phi\phi}}{U},\quad\beta = \frac{1}{8\pi G}\left(\frac{\Gamma_{\phi}\,U_\phi}{\Gamma\,U}\right),
\end{equation}
the condition in Eq.~(\ref{slow_roll_conditions}) is met if the slow-roll parameters satisfy \cite{visinelli_NWI, taylor_berera, hall, moss_xiong, moss}
\begin{equation} \label{slow_roll}
\epsilon \ll 1+\frac{\Gamma}{3H},\quad |\eta| \ll 1+\frac{\Gamma}{3H},\quad |\beta| \ll 1+ \frac{\Gamma}{3H}.
\end{equation}

We now discuss the evolution of perturbations during the slow-roll regime. In the following, we take into account the effects from both quantum and thermal fluctuations, the latter being the predominant source for perturbations in WI scenarios. For this, we introduce a stochastic noise $\xi({\bf x},t)$ in the source term for $\delta \rho_\phi$, according to the Schwinger-Keldysh approach to describe the quantum mechanical evolution of a system in a non-equilibrium state~\cite{schwinger, keldysh, rammer}. This term has been discussed by Calzetta and Hu~\cite{calzetta_hu} in the context of the Boltzmann equation, and by Berera and Fang~\cite{berera_fang} in the setting of WI. Instead of $\delta Q_\phi$ in Eq.~(\ref{cold_perturbations}), we thus consider the source term 
\begin{equation} \label{stochastic_source}
\delta Q_\phi = -\Gamma\,\left(\delta p_\phi + \delta \rho_\phi\right) - \dot{\bar{\phi}}\,\Gamma_{\rm eff}\,\xi_q,
\end{equation}
where $\xi_q$ is the Fourier transform of the stochastic noise, which follows the statistical average
\begin{equation} \label{condition_stochastic1}
\displaystyle \langle \xi_q\,\xi_{q'}\rangle = a(t)^{-3}\,\delta^{(3)}({\bf q + q'})\,\delta(t-t'),
\end{equation}
and we have defined the parameter~\cite{moss}
\begin{equation}
\Gamma_{\rm eff} = \sqrt{2\,\left[\Gamma + H\right]\,T},
\end{equation}
where the temperature of the radiation bath is defined through $\bar{\rho}_r$ in Eq.~(\ref{energy_pressure}). Stochastic perturbations may arise from microscopical models for radiation, as discussed for example in Refs.~\cite{bastero12b, bastero14, bastero13} in the context of supersymmetric models. In this work, we decided to treat the radiation density $\bar{\rho}_r$ as a fluid, and to introduce the stochastic source $\xi_q$ in a phenomenological way, without making assumptions on the underlying model for radiation. This will be the subject of a future work, along the line of Ref.~\cite{visinelli_NWI}, in which perturbations in models of axion-like particles are discussed.

With the addition of the stochastic source perturbations in Eq.~\eqref{stochastic_source}, the equation for $\delta \phi$ in Eq.~\eqref{perturbation_eq_phi} modifies as
\begin{equation} \label{perturbation_eq_stochastic}
\delta \ddot{\phi}_q + \left(3H +\Gamma\right)\,\ddfq + \left(U_{\phi\phi} + \frac{q^2}{a^2}\right)\,\dfq - 4\,\dot{\bar{\phi}}\,\dot{\Psi}_q + \left(2\,U_\phi + \Gamma\,\dot{\bar{\phi}}\right)\,\Psi_q = -\Gamma_{\rm eff}\,\xi_q.
\end{equation}
We have checked that this expression is recovered also by using the results in Ref.~\cite{bastero1}, once set to zero the perturbations on the dissipation term $\delta \Gamma_q = 0$ which we do not include.

We assume that, during the slow-roll regime, the gravitational field perturbation varies slowly as \cite{polarski}
\begin{equation}
\dot{\Psi}  \ll H\,\Psi, \quad
\end{equation}
so that, using Eq.~(\ref{eq_motion_slow_roll}), perturbations for the inflaton field in Eq.~(\ref{perturbation_eq_stochastic}) are given by
\begin{equation} \label{perturbation_slow_roll}
\delta \ddot{\phi}_q + \left(3H +\Gamma\right)\,\ddfq + \left[U_{\phi\phi} + \frac{q^2}{a^2} - \frac{2H\,(3H + \Gamma)(6H+\Gamma)}{\Gamma}\,\frac{\bar{\rho}_r}{U}\right]\,\dfq \simeq -\Gamma_{\rm eff}\,\xi_q.
\end{equation}
We redefine the latter term in the square brackets as
\begin{equation}
 - \frac{2H\,(3H + \Gamma)(6H+\Gamma)}{\Gamma}\,\frac{\bar{\rho}_r}{U} = \kappa\,\frac{U_\phi^2}{2U},
\end{equation}
where we have parametrizes the strength of $\Gamma$ with respect to $H$, as $\kappa = 1$ when $H \gg \Gamma$ and $\kappa = 1/2$ when $H \ll \Gamma$. With this substitution, the first line of Eq.~(\ref{perturbation_slow_roll}) gives the dissipation equation presented in Ref.~\cite{moss}, except for the extra term containing $U_{\phi\phi}$ and $U_\phi^2$,
\begin{equation}
\delta\ddot{\phi}_q + \left(3H +\Gamma\right)\,\delta \dot{\phi}_q + \left(U_{\phi\phi} + \kappa \,\frac{U_\phi^2}{2U} + \frac{q^2}{a^2}\right)\,\delta\phi_q \simeq -\Gamma_{\rm eff}\,\xi_q,
\label{perturbation_slow_roll1}
\end{equation}
In order to solve for $\phi_q$, we switch to the variable
\begin{equation}
z = \frac{q}{H\,a(t)},
\end{equation}
and, using the slow-roll Eqs.~(\ref{slow_roll_parameters}) in the form $U_\phi^2 = 3H^2\,U\,\epsilon/2$ and $U_{\phi\phi} = 3H^2\,\eta$, we define
\begin{equation}
\nu = \frac{3}{2}\,(1+R),\quad \hbox{and}\quad \mu = \sqrt{\nu^2 - 3\,(\kappa\,\epsilon + \eta)}.
\end{equation}
Eq.~(\ref{perturbation_slow_roll1}) is rewritten as
\begin{equation}
\begin{array}{l}
\displaystyle \frac{d^2\,\delta\phi}{dz^2} + \frac{1-2\nu}{z}\,\frac{d\,\delta \phi}{dz} + \left[1+\frac{3\,(\kappa\,\epsilon + \eta)}{z^2}\right]\,\delta \phi = \frac{\left[2(\Gamma+H)\,T\right]^{1/2}}{H^2\,z^2}\,\xi_q,
\end{array}
\label{diff_eq_perturbation_phi_moss1}
\end{equation}
which is a second order stochastic differential equation, with solution for the field and its first derivative given respectively by
\begin{equation}
\begin{array}{l}
\displaystyle \delta \phi_q(z) = z^\nu\,\left(\delta\phi_1\,J_\mu(z) + \delta\phi_2\,Y_\mu(z)\right) -\frac{\pi}{2}\,\frac{\left[2(\Gamma+H)\,T\right]^{1/2}}{H^2}\,\int_z^{+\infty}\,G^{(1)}_\mu(z,z')\,\left(\frac{z}{z'}\right)^\nu\,\xi_q(z')\,\frac{dz'}{z'},\\
\displaystyle\\
\displaystyle \delta \phi'_q(z) = \delta\phi'_1+\delta\phi'_2 -\frac{\pi}{2}\,\frac{\left[2(\Gamma+H)\,T\right]^{1/2}}{H^2}\,\int_z^{+\infty}\,G^{(2)}_\mu(z,z')\,\left(\frac{z}{z'}\right)^\nu\,\xi_q(z')\,\frac{dz'}{z'}.
\end{array}
\end{equation}
Here, $J_\mu(z)$ and $Y_\mu(z)$ are the Bessel functions of respectively the first and second kind of order $\mu$, $\delta\phi_2$ and $\delta\phi_2$ are arbitrary constants, and we have introduced the functions
\begin{equation}
\begin{array}{l}
G^{(1)}_\mu(z,z') = J_\mu(z)\,Y_\mu(z')-J_\mu(z')\,Y_\mu(z),\\
G^{(2)}_\mu(z,z') = J_{\mu-1}(z)\,Y_\mu(z')-J_\mu(z')\,Y_{\mu-1}(z).
\end{array}
\end{equation}
This solution generalizes the result presented in Ref.~\cite{moss} to the case $\mu \neq \nu$, and it is valid whenever the slow-roll parameters can be considered constant.

\section{Conclusions} \label{Discussion and conclusions}

We shall summarize the main points of this paper. In Sec.~\ref{Equations for the background fields}, we have revised the Boltzmann equations for an inflaton field coupled to radiation, and we have solved the system of equation numerically in the case of a quartic inflaton potential $U(\phi)$, see Eq.~\eqref{potential}. In Sec.~\ref{Evolution of perturbations during inflation}, we have applied the interacting fluids formalism, based on the framework given in Refs.~\cite{kodama, mukhanov1992} and described in Refs.~\cite{malik, malik1}, to derive the differential equations governing the cosmological fluctuations arising during the inflationary period in Eqs.~\eqref{perturbation_eq_phi}-\eqref{perturbation_eq_psi}. These equations have been solved numerically for a quartic inflaton potential $U(\phi)$. We give an analytical expression for $\delta \phi$ around the reheating period in Eq.~(\ref{perturbation_eq_X}), which approximates the frequency of oscillation obtained with the numerical solution. 

We have considered the effects of a viscosity term $\eta_0$ in Sec.~\ref{dissipative_terms}, following the result of other work in the literature~\cite{delcampo, bastero12a, bastero1}. We have obtained that a constant viscosity term predicts a constant value for the energy density of the radiation field in the post-inflationary period, see Eq.~\eqref{viscosity_radiation_rho}, and affects perturbations in the energy density and velocity of the fluid by damping oscillations. These occurrences can be avoided in theories which predict a time-depending $\eta_0$, as we have shown in Fig.~\ref{figure:eta} for a specific function of time. These results will be further explored in a future work on realistic microscopic models of the radiation field. In conclusion, we discuss perturbations in the presence of viscosities in Eqs.~\eqref{perturbation_eq_phi_viscosity1}-\eqref{perturbation_field_viscosity1}, and we show numerical results in Fig.~\ref{figure:eta}.

In Sec.~\ref{Stochastic perturbations}, we added a stochastic source to the expression describing $\delta \phi_q$, which is the approach usually considered in warm inflation scenarios. In the slow-roll regime, the equation for the inflaton perturbations can be solved exactly, including the stochastic term. The solution extends the result obtained in Ref.~\cite{moss} to the case in which the slow-roll parameters are not neglected in the differential equation.

\begin{acknowledgments}
The author would like to thank Paolo Gondolo and Xuefang Sui for useful discussions on cosmological perturbations.
\end{acknowledgments}

\end{document}